\documentstyle[aps,prbbib,epsf]{revtex}

\begin{document}

\title{
The Transition State in Magnetization Reversal
}

\author{
G. Brown$^{1,2}$, 
M. A. Novotny$^{3}$,
Per Arne Rikvold$^{2,4}$
}

\address{
$^{1}$Center for Computational Sciences, 
      Oak Ridge National Laboratory, 
      Oak Ridge, TN 37831\\
$^{2}$CSIT,
      Florida State University,
      Tallahassee, FL 32306\\
$^{3}$Department of Physics and Astronomy and ERC,
      Mississippi State University,
      Mississippi State, MS 39762\\
$^{4}$Department of Physics and MARTECH, 
      Florida State University,
      Tallahassee, FL 32306\\
}

\date{\today}
\maketitle

\begin{abstract}
We consider a magnet with uniaxial anisotropy in an external magnetic
field along the anisotropy direction, but with a field magnitude
smaller than the coercive field. There are three representative
magnetization configurations corresponding to three extrema of the
free energy.  The equilibrium and metastable configurations, which are
magnetized approximately parallel and antiparallel to the applied
field, respectively, both correspond to local free-energy minima.  The
third extremum configuration is the saddle point separating these
minima.  It is also called the transition state for magnetization
reversal. The free-energy difference between the metastable and
transition-state configurations determines the thermal stability of
the magnet. However, it is difficult to determine the location of the
transition state in both experiments and numerical simulations. Here
it is shown that the computational Projective Dynamics method, applied
to the time dependence of the total magnetization, can be used to
determine the transition state. From large-scale micromagnetic
simulations of a simple model of magnetic nanowires with no
crystalline anisotropy, the magnetization associated with the
transition state is found to be linearly dependent on temperature, and
the free-energy barrier is found to be dominated by the entropic
contribution at reasonable temperatures and external fields.  The
effect of including crystalline anisotropy is also discussed. Finally,
the influence of the spin precession on the transition state is
determined by comparison of the micromagnetic simulations to kinetic
Monte Carlo simulations of precession-free (overdamped) dynamics.
\end{abstract}

\vskip 0.15in

Since magnetic storage depends on the ability to quickly assign a
magnetic orientation to a specific region of a medium, in which
spontaneous reorientation is suppressed, understanding the process of
magnetization reversal is technologically important. In magnetic
fields smaller than the coercive value, the two states of the magnet
important to information storage are free-energy minima separated by a
free-energy barrier. The minimum corresponding to a magnetization
parallel to the field is truly stable, while the minimum corresponding
to the antiparallel magnetization is higher in energy and is therefore
metastable. The free-energy maximum along the most likely path between
the minima is a saddle point. To effectively engineer processes such
as hybrid recording, \cite{Ruigrok} which use lower-than-coercive
applied fields to assign magnetic orientations in high-coercivity
magnetic materials, a thorough understanding of the magnetization
reversal is needed. Properties of the free energy near the minima and
saddle point often control the nonequilibrium dynamics of a
system. For example, the free-energy difference between a minimum and
the saddle point, the barrier height, is a measure of the stability of
the magnetization associated with that minimum, while the curvature of
the free energy near a minimum is a measure of the ``attempt
frequency'' of escapes over the barrier. While the free-energy minima
can be easily located as local maxima in histograms of the state of
the system, the saddle point has proven much harder to measure. In
this paper we present a method based on the Projective Dynamics method
\cite{KOLESIK,NOVOTNY,MITCHELL} for determining the saddle point in
magnetization reversal of nanoscale magnets, and then investigate the
free-energy barrier associated with the process.

The particular magnetic system considered in this paper is based on
iron nanomagnets fabricated using STM-assisted chemical vapor
deposition. \cite{WIRTH} These magnetic pillars are modeled by a
one-dimensional array of $17$ spins along the $z$ axis, each a unit
vector ${{\hat S}_i}$ representing the local orientation of the
magnetization.  The total energy of the pillar is given by
\begin{equation}
E = -\frac{J}{2} \sum_{i,j} {\hat S}_i \cdot {\hat S}_j
    -\frac{D}{2} \sum_{i,j \ne i} {\hat S}^t_i {\bf T}_{i,j} {\hat S}_j 
    -K \sum_{i} (S_{i,z})^2 + B \sum_{i} S_{i,z}
\;,
\end{equation}
where $J$ is the exchange energy, $K$ is the uniaxial anisotropy
energy, $B$ is the strength of the external field oriented parallel to
$-{\hat z}$, $D$ is the strength of the dipole-dipole interactions in
energy units, and ${\bf T}_{i,j}=(3{\hat z}{\hat z}^t-{\bf
1})/|j-i|^3$ is the dimensionless dipole-dipole interaction
tensor. The sum for the exchange energies is over nearest neighbors
and includes a self interaction so that the exchange energy (and the
associated effective field) is zero for parallel neighbors. For the
$17$-spin iron nanopillars, discussed previously in Ref.~[6], the
dimensionless energy corresponds to $J$$=$$1.6 \times 10^{-12}$ erg,
$D$$=$$4.1\times 10^{-12}$ erg, and $B$$=$$1.9 \times 10^{-12}$ erg
for a field of $1000$ Oe. Dipole-dipole interactions along the pillar
provide a strong uniaxial anisotropy, even for $K$$=$$0$.

For micromagnetic dynamics, each vector precesses around a local field
${{\vec H}_i} = -{dE}/{d{\hat S}_i}$ according to the
Landau-Lifshitz-Gilbert (LLG) equation \cite{BROWN63,AHARONI}
\begin{equation}
\frac{ {d} {{\hat S}_i}}
     { {d} {t} }
 =
   \frac{ \gamma_0 }
        { 1+\alpha^2 }
   {{\hat S}_i}
 \times
 \left[
   {{\vec H}_i}
  -{\alpha} {{\hat S}_i} \times {{\vec H}_i}
 \right]
\;,
\end{equation}
where the scaled electron gyromagnetic ratio $\gamma_0 = 9.26\times10^{21}$
Hz/erg for this system. The phenomenological damping parameter
$\alpha=0.1$ was chosen to give underdamped dynamics.

Previously, \cite{UGA02} we have shown that the Projective Dynamics
method \cite{KOLESIK,NOVOTNY,MITCHELL} can be used to locate the
saddle point. The computational Projective Dynamics method involves
projecting the original description of the dynamics in terms of a
large number of variables onto a completely stochastic description in
terms of one or two variables. In this paper, Eq~(1) has $34$
independent variables which are projected onto a single, slowly
varying variable: the $z$-component of the total magnetization, $M_z.$
Furthermore, the values of $M_z$ are binned, and the transition rates
between bins are measured. For instance, the probability that $M_z$
starts a given time interval in one bin and ends the interval in the
adjacent bin corresponding to smaller values of $M_z$ is called the
``growth'' probability, $P_{\rm grow}$, since it corresponds to an
increase in the volume of stable magnetization. The probability of
moving to the opposite neighboring bin is $P_{\rm shrink}.$

The measured $P_{\rm grow}$ and $P_{\rm shrink}$ for an applied field
of $H$$=$$1000$ Oe, a temperature of $T$$=$$10$ K, and $K$$=$$0$ are
shown in the inset of Fig.~1 as the solid and broken curves,
respectively.  For magnetizations with $P_{\rm grow}$$>$$P_{\rm
shrink}$, on average the magnetization decreases with time, while it
increases, on average, for magnetizations with $P_{\rm
shrink}$$>$$P_{\rm grow}$. Where $P_{\rm grow}$$=$$P_{\rm shrink}$ the
average rate of change is zero, and a local minimum or maximum in the
free energy occurs. The right-most crossing of $P_{\rm grow}$ and
$P_{\rm shrink}$ is the free-energy minimum corresponding to the
metastable magnetization antiparallel to the applied field. The
crossing associated with the free-energy minimum of the stable
magnetization (near $M_z$$=$$-1$) is not shown. The left-most crossing
shown in the figure corresponds to the free-energy saddle point
separating the metastable and stable wells.

The value of $M_z$ at the saddle point for $H$$=$$1000$ Oe and
$K$$=$$0$ is shown in Fig.~1 vs $T$ as circles with error bars. These
values were estimated from the intersection of lines fit to $P_{\rm
grow}$ and $P_{\rm shrink}$ in the region near the crossing. The
measurement errors were taken to be $\pm 0.01,$ based on the size of
the region where $P_{\rm grow}$ $\approx$ $P_{\rm shrink}$. For low
$T$, the estimated value of $M_z$ depends almost linearly on
temperature.  Good agreement exists between the estimated values and
the line of best fit, which was determined from the data for
$T$$\le$$100$ K and is shown as the solid line in Fig.~1.

The Projective Dynamics method can also be used to estimate the
free-energy barrier for the reversal process. In equilibrium, the
total time spent in the $i$th $M_z$ bin, $h(i)$, is related to the
growing and shrinking probabilities by \cite{NOVOTNY,MITCHELL}
\begin{equation}
h(i) = \left[1+P_{\rm shrink}(i-1)h(i-1)\right]/P_{\rm grow}(i), 
\qquad h(1)=1/P_{\rm grow}(1)
\;.
\end{equation}
The time $h(i)$ is also related to the free energy by the Boltzmann
factor $h(i) \propto \exp(-F(i)/k_BT).$ Taking the bin containing the
saddle point to be $i_{\rm s}$, and that containing the metastable minimum
to be $i_{\rm m},$ the free-energy barrier is $\Delta F = k_B T
\ln[h(i_{\rm m})/h(i_{\rm s})].$ The measured $\Delta F$ for 
$H$$=$$1000$ Oe and $K$$=$$0$ are shown vs temperature as the circles
in Fig.~2. The measured $\Delta F$ is not constant, as would be
expected in the commonly assumed Arrhenius model of thermally
activated processes. What is even more striking is that $\Delta E$,
the difference in average energy between the bin containing the saddle
point and that containing the metastable minimum as calculated by
Eq.~(1), is always negative. (The absolute value of $\Delta E$ is of
the same order of magnitude as $\Delta F,$ but generally lies within
one standard deviation of zero.)  This is consistent with $E$ vs $M_z$
shown in the inset of Fig.~2 (along with the contribution from the
various terms) for one reversal at $H$$=$$1000$ Oe, $T$$=$$10$ K, and
$K$$=$$0$. The total energy is seen to be nearly constant for
$M_z$$>$$0.80,$ which is the region around the barrier. Since the free
energy is the sum of $E$ and $-TS$, where $S$ is the entropy, it is
reasonable to conclude that the dominant contribution to the positive
free-energy barrier is a decrease in entropy going from the metastable
minimum to the saddle point.

It should be noted that $\Delta E$ is measured from the total energy
averaged in bins of $M_z$, and it is possible that the saddle point is
associated with extreme values of $E$ within the bin. In that case,
even though $\Delta E$ measured from the average energy of the bin is
negative, it could be possible that the $\Delta E$ associated with
the reversal is positive. However, no support for extreme values of
$E$ is seen in $E$ vs $M_z$ or $E$ vs $t$ for individual reversals
under a variety of conditions.

The effect of a weaker applied field was also investigated. For
$H$$=$$900$ Oe and $K$$=$$0$ results are shown in both figures as
squares. There is a small decrease in $M_z$ at the saddle point, and a
significant increase in the free-energy barrier. These results are
what one normally expects when the field is lowered, and the
difference in energies between the metastable and stable
magnetizations decreases. Results for $H$$=$$1000$ Oe and an
anisotropy energy of $K$$=$$1.9 \times 10^{-13}$ erg are shown in the
figures as diamonds. The broken line in Fig.~1 is a best fit, which
shows that $M_z$ at the saddle point is much less temperature
dependent, than in the $K$$=$$0$ case. The free-energy barrier is
increased by increasing $K$. Although the estimates of $\Delta F$ are
more uncertain for $K$$=$$1.9 \times 10^{-13}$ erg (a consequence of
decreased sampling due to the higher barrier), and data are available
only for $T$$\ge$$40$ K, the data indicate that the temperature
dependence of $\Delta F$ may be suppressed by nonzero $K$. The measured
$\Delta E$ are negative, even for this value of $K$.

The reversal process was also simulated with a precession-free kinetic
Monte Carlo (MC) method, which corresponds to the overdamped limit of
the LLG equation, \cite{Nowack} to estimate the effect of the
underdamped precession of the spins. In this dynamic, one spin is
selected at random and is tentatively displaced by a step uniformly
distributed on a sphere of radius $R$$=$$0.005$. Using Eq.~(1) to
calculate the energies, the displacement is accepted or rejected using
the Metropolis method. The spins are normalized after each move to
conserve their length. Monte Carlo simulations were performed only
for $H$$=$$1000$ Oe and $K$$=$$0.$ The $M_z$ of the saddle point, the
triangles in Fig.~1, are larger than those found using the LLG
dynamics, but the dependence on temperature appears to be linear. The
free-energy barrier is smaller than that measured for the LLG dynamics
at high temperature, but the difference decreases as the temperature
is decreased. Even for precession-free MC dynamics, all measured
$\Delta E$ are negative.

In summary, details of the magnetization reversal in underdamped,
nanoscale magnets have been investigated with Projective Dynamics. The
magnetization at the saddle point has been measured and found to
decrease linearly with temperature for $T$$<$$100$ K. In contrast to
the Arrhenius model, which assumes a temperature-independent barrier,
the measured free-energy difference between the saddle point and the
metastable magnetization is found to be temperature dependent. For the
physically reasonable parameters considered here, the difference in
the average energy between the saddle point and the metastable
magnetization is negative, while the difference in the free energy is
positive. From this we conclude that the free-energy barrier under
these conditions is due to reduced entropy at the saddle
point. Although the temperature dependence of the barrier is weaker,
Monte Carlo simulations of overdamped dynamics show the same dominance
of entropic contributions to the free-energy barrier
 
This work was supported by the Laboratory Directed Research and
Development Program of Oak Ridge National Laboratory, managed by
UT-Battelle, LLC for the U.S. Department of Energy under Contract
No. DE-AC05-00OR22725, by the Ames Laboratory, which is operated
for the U.S. Department of Energy by Iowa State University under
Contract No. W-7405-82, by the National Science Foundation
under Grant 0120310, and by Florida State University through CSIT
and MARTECH.

\newpage
\begin{figure}[tb]
\centerline { \hbox{\epsfxsize 3in \epsfbox{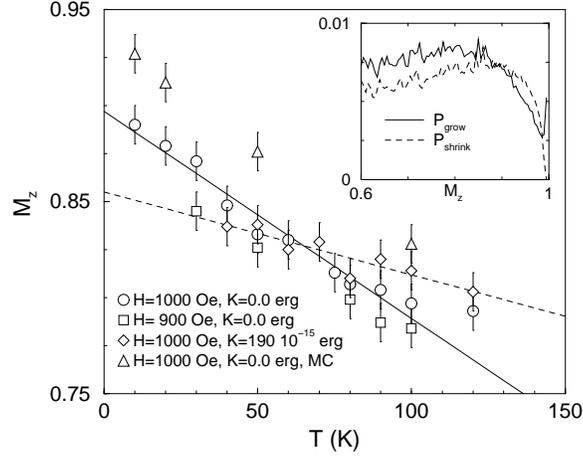} } }
\caption[]{Saddle-point magnetization along the pillar, $M_z$, estimated from
$P_{\rm grow}$$=$$P_{\rm shrink}$, shown vs temperature, $T$, under
different conditions. The solid line is a best fit to the results for
$T$$\le$$100$ K at $H$$=$$1000$ Oe and $K$$=$$0$, and the broken line
is a best fit for $H$$=$$1000$ Oe and $K$$=$$190 \times 10^{-15}$
erg. The data for a weaker field, $H$$=$$900$ Oe, and over-damped
dynamics also change approximately linearly with temperature. The
probabilities $P_{\rm grow}$ and $P_{\rm shrink}$ for $H$$=$$1000$ Oe,
$T$$=$$10$ K, and $K$$=$$0$ are shown in the inset. The right-most
crossing occurs at the metastable minimum, and the left-most crossing shown
occurs at the saddle point.}
\end{figure}

\begin{figure}[tb]
\centerline{\epsfxsize 3in \epsfbox{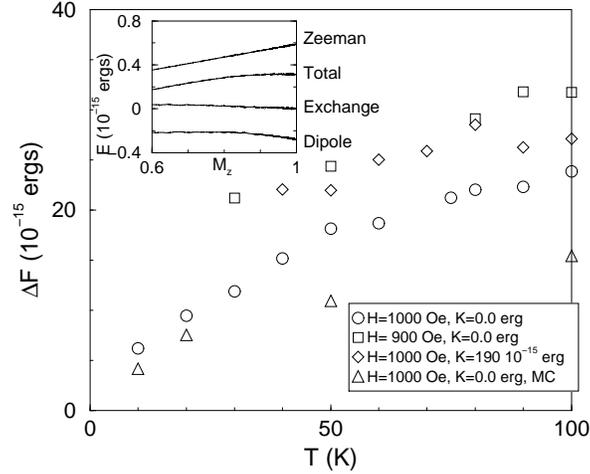} }
\caption[]{Free-energy barrier associated with magnetization reversal,
$\Delta F$, measured from the residence time, $h$, shown vs
temperature $T$. The measured switching times for various conditions
varied from $7.5$ ns ($T$$=$$90$ K, $H$$=$$1000$ Oe, $K$$=$$0$) to
$143.2$ ns ($T$$=$$10$ K, $H$$=$$1000$ Oe, Monte Carlo). All energy
differences, $\Delta E$, are negative (not shown). The inset shows the
energy, defined in Eq.~(1), vs $M_z$ for one reversal event at
$H$$=$$1000$ Oe, $T$$=$$10$ K, and $K$$=$$0$. The curve is parametric
with respect to the simulation time, and it is not single valued near
the metastable magnetization.  }
\end{figure}

\end{document}